# Evaluation of deep convolutional neural networks in classifying human embryo images based on their morphological quality

Prudhvi Thirumalaraju[1]†, Manoj Kumar Kanakasabapathy[1]†, Charles L Bormann[2,3]†, Raghav Gupta[1], Rohan Pooniwala[1], Hemanth Kandula[1], Irene Souter[2,3], Irene Dimitriadis[2,3], Hadi Shafiee[1,3]*

[1] Division of Engineering in Medicine, Department of Medicine, Brigham and Women's Hospital, Harvard Medical School, Boston, Massachusetts, United States of America

[2] Division of Reproductive Endocrinology and Infertility, Department of Obstetrics & Gynecology, Massachusetts General Hospital, Harvard Medical School, Boston, Massachusetts, United States of America

[3] Department of Medicine, Harvard Medical School, Boston, Massachusetts, United States of America

*Corresponding author: Hadi Shafiee

E-mail: hshafiee@bwh.harvard.edu
† These authors contributed equally to this work

## Abstract

A critical factor that influences the success of an in-vitro fertilization (IVF) procedure is the quality of the transferred embryo. Embryo morphology assessments, conventionally performed through manual microscopic analysis suffer from disparities in practice, selection criteria, and subjectivity due to the experience of the embryologist. Convolutional neural networks (CNNs) are powerful, promising algorithms with significant potential for accurate classifications across many object categories. Network architectures and hyper-parameters affect the efficiency of CNNs for any given task. Here, we evaluate multi-layered CNNs developed from scratch and popular deep-learning architectures such as Inception v3, ResNET, Inception-ResNET-v2, and Xception in differentiating between embryos based on their morphological quality at 113 hours post insemination (hpi). Xception performed the best in differentiating between the embryos based on their morphological quality.

# 1. Introduction

Infertility is an underestimated healthcare problem that affects over 48 million couples globally and is a cause of distress, depression, and discrimination (Mascarenhas et al., 2012; Turchi, 2015). Although assisted reproductive technologies (ART) such as in-vitro fertilization (IVF) have alleviated the disease burden to an extent, it has been inefficient with an average success rate of approximately 30% reported in 2015 in the US (CDC, 2015). IVF remains an expensive solution costing $7000 - $20,000 per ART cycle in the US, most of which is not covered by insurance (Birenbaum-Carmeli, 2004; CDC, 2015; Toner, 2002) with many patients requiring multiple cycles to achieve a successful pregnancy. Multiple factors such as maternal age, medical diagnosis, gamete and embryo quality, and endometrium receptivity determine the success of ART cycles (Barash et al., 2017; Demko et al., 2016; Einarsson et al., 2017; Erenus et al., 1991; Hill et al., 1989; Osman et al., 2015; Paulson et al., 1990). However, non-invasive selection of the highest available quality from a patient's cohort of embryos (top-quality embryo) for transfer, remains one of the most important factors in achieving successful ART outcomes, yet this critical step remains a significant challenge (Barash et al., 2017; Conaghan et al., 2013; Filho et al., 2010; Machtinger and Racowsky, 2013; Racowsky et al., 2015; Vaegter et al., 2017; Wong et al., 2013).

Embryo transfers are performed at the cleavage or the blastocyst stage of development. Embryos are at the cleavage stage 2-3 days after fertilization and reach the blastocyst stage 5-7 days after fertilization. Traditional methods of embryo selection rely on visual embryo morphological assessment and are highly practice-dependent and subjective. Emulating the skill of highly trained embryologists in efficient embryo assessment in a fully automated system is a major unmet challenge in all of the previous work done in embryo computer-aided assessments due to focus on measuring specific expert-defined parameters such as zona pellucida thickness variation, number of blastomeres, degree of cell symmetry and cytoplasmic fragmentation, etc. (Rocha et al., 2017a; Rocha et al., 2017b). Computer vision methods for embryo assessment are semi-automated, limited to measuring specific parameters providing metrics that require further analysis by embryologists and strictly controlled imaging systems (Filho et al., 2010). Previous attempts in developing systems using traditional machine-learning approaches require intensive image preprocessing followed by human-directed segmentation of embryo features for classification (Matos et al., 2014; Rocha et al., 2017a; Rocha et al., 2017b). Owing to the dependency of these approaches on image processing and segmentation, such methods suffer from the same limitations as computer vision techniques.

Convolutional neural networks (CNN), which work on the principles of representation learning, are excellent candidates for this application, have already received significant attention from the medical community in evaluating, through embryo morphology, the embryo quality and implantation potential, and for other applications for clinical IVF practices such as quality control of systems and embryologists (Dimitriadis et al., 2019a; Dimitriadis et al., 2019b; Hariton et al., 2019; Kanakasabapathy et al., 2019a; Kanakasabapathy et al., 2019b; Kanakasabapathy et al., 2019c; Khosravi et al., 2019; Thirumalaraju et al., 2019a; Thirumalaraju et al., 2019b; Thirumalaraju et al., 2019c; Tran et al., 2019). However, all of these studies provide limited information on the neural networks themselves and the effect of hyper-parameters on the task of embryo morphological assessments. Different architectures and hyper parameters achieve varying performances on the same task and are non-universal. Therefore, the primary goal of this study is to evaluate popular CNN approaches using a dataset of day 5 embryo (113 hours post insemination) images in classifying embryos based on their morphological quality. Day 5 embryos were used for the network evaluation studies given their importance to field of embryology and since most published studies are focused this day of development (Dimitriadis et al., 2019b; Hariton et al., 2019; Kanakasabapathy et al., 2019a; Kanakasabapathy et al., 2019b; Kanakasabapathy et al., 2019c; Khosravi et al., 2019; Thirumalaraju et al., 2019b; Thirumalaraju et al., 2019c).

Embryos with normal fertilization were evaluated based on their morphology at 113 hours post insemination (hpi) (Fig 1). At the blastocyst stage (at 113 hpi), embryos are conventionally graded through 83 classes of blastocysts based on the combinations of (i) the degree of blastocoel expansion (grades 1-6), (ii) inner cell mass quality (grades 1-4), and (iii) trophectoderm quality (grades 1-4) along with 3 classes of non-blastocysts. For the CNN classification algorithm, the grading system was simplified to encompass all 86 classes within a 2-level hierarchy of training and inference classes (Fig 1). Thus, the embryos were evaluated at 113 hpi stage using multilayered CNNs (5 to 43 layers), Inception v3 (Szegedy et al., 2015), ResNET (He et al., 2015), Inception-ResNET-v2 (Szegedy et al., 2016), and Xception (Chollet, 2016). The two major categories of non-blastocysts and blastocysts included the training classes 1, 2, and 3, 4, 5, respectively. Using a retrospective dataset comprising of 2,440 embryos, the deep CNN models were trained and tested to primarily classify between two classes (non-blastocysts and blastocysts) using images of embryos captured at 113 hpi. The best performing model was then used to evaluate an independent test set of 742 embryos in differentiating blastocysts based on their morphological quality. In addition, we also tested the networks (Inception-v3, ResNET, Inception ResNET, multilayer CNN, and Xception) in evaluating embryo images with a domain shift.

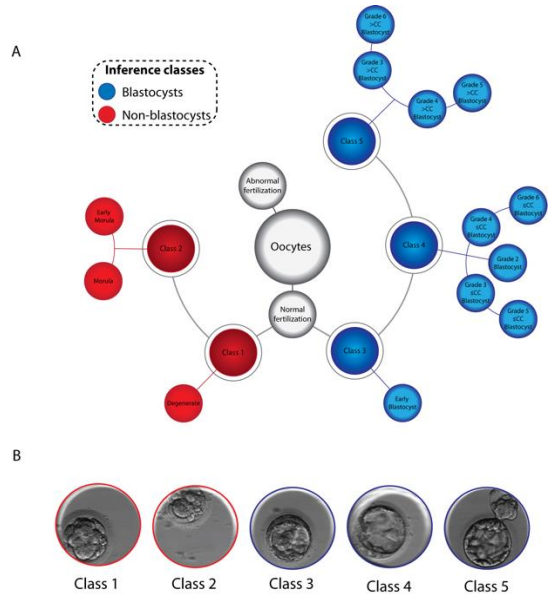

**Fig 1. Embryo hierarchy used by the neural network.** (A) Following insemination, pronuclear stage embryos are categorized into two classes and based on their 113-hours morphologies are sorted into 2 major classes (blastocysts and non-blastocysts) subdivided into 5 classes. Embryos with abnormal fertilization (non-2PN) were not tracked further and thus were not considered in 70 hpi and 113 hpi assessments. Classes 1 and 2 were composed of non-blastocysts and classes 3, 4, and 5 were composed of blastocysts. Class 5 composed of blastocysts that met the clinical criteria for cryopreservation. (B) Representative images for each class of embryo.

## 2. Materials and methods

### 2.1 Data collection and preparation

Data was collected at the Massachusetts General Hospital (MGH) fertility center in Boston, Massachusetts. We used 3,469 recorded videos of embryos collected from 543 patients under an institutional review board approval (IRB#2017P001339; IRB#2019P002392). The retrospective image data used for this study were collected as part of routine clinical practice using an Embryoscope time-lapse system (Vitrolife). These instruments use Hoffman modulated contrast optics with 20x objective to image each embryo. Images are acquired at a resolution of 1280 x 1024 pixels every 10 mins at 7 focal planes, to generate videos. Videos were fragmented to extract the frames at a single focal plane and linked to specific a time point (113 hpi) using a custom python script, which made use of the OpenCV and Tesseract libraries. Machine-generated timestamps available on each frame of the video was used to identify the images associated with 113 hpi. All embryos used in the study were annotated using images from the fixed time-points by senior-level embryologists with a minimum of 5 years of human IVF training. Out-of-focus images were included in the datasets and used for both testing and training. Only images of embryos that were completely non-discernable were removed as part of the data cleaning procedure.

### 2.2 Data organization and hierarchical structuring

Embryo images collected at 113 hpi were separated prior to evaluation. Only embryos with normal fertilization were used for evaluations. The embryo images at 113 hpi time points were categorized between training classes 1 through 5 (Fig. 1). The embryo class categorizations were based on the embryos' developmental state achieved by 113 hpi. Class 1 comprised of degenerated and arrested embryos, which did not begin compaction while class 2 comprised of embryos that were at the morula stage at 113 hpi. Classes 1 and 2 together formed the inference class of 'non-blastocysts'. Class 3 comprised of embryos exhibiting features of an early blastocyst such as the presence of a blastocoel cavity and a thick zona pellucida with lack of overall embryo expansion. Class 4 was made up of embryos, which were blastocysts with blastocoel cavities occupying over half of the embryo volume and possessed either poor inner cell mass (ICM) or poor trophectoderm (TE). These embryos were overall considered to fall below 113 hpi cryopreservation quality criteria based on the MGH fertility center guidelines (> 3CC), where 3 represents the degree of expansion (range 1-6) and C represents the quality of ICM and TE (range A-D), respectively (Table 1). Class 5 on the other hand comprised of all embryos, which met cryopreservation criteria and included full blastocysts to hatched blastocysts (Table 1). Classes 3, 4, and 5 together formed the inference class of 'blastocysts' that was used in this study.

| Day 5/6 Stage (>113 hpi) | Score | Class | Description |
|---|---|---|---|
| Degenerate or Arrested | D | 1 | Embryo failed to develop to at least the morula stage |
| Morula | M-A | 2 | More than 50% of the embryo has undergone compaction; no ICM or TE cells evident |
| Morula | M-B | 2 | Incomplete compaction (less than 50% compaction) |
| Early Blastocyst | 1* | 3 | Blastocoele less than half the volume of the embryo, little or no expansion in overall size; ZP thick (1A = good quality; 1B = moderate quality, 1C = poor quality) |
| Blastocyst | 2 | 4 | Blastocoele more than half the volume of the embryo, some expansion in overall size; ZP beginning to thin |
| Full Blastocyst | 3 | 4 or 5** | Blastocoele completely filling embryo; ZP not completely thinned |
| Expanded Blastocyst | 4 | 4 or 5** | Blastocoele completely filling embryo; fully expanded embryo and ZP very thin |
| Hatching Blastocyst | 5 | 5 | Hatching blastocyst, TE starting to herniate through the ZP |
| Hatched Blastocyst | 6 | 5 | Blastocyst completely hatched (i.e. completely out of the ZP) |
| **ICM Grade** | | | **Description** |
| A | | | ICM prominent & easily discernible with many cells, and cells compacted and tightly adhered together |
| B | | | ICM discernible but with fewer cells, and loosely adherent together |
| C | | | Very few cells visible, either compacted or loose, may be difficult to distinguish completely from TE |
| D | | | No ICM cells discernible in any focal plane or ICM cells appear degenerate or necrotic |
| **Trophectoderm Grade** | | | **Description** |
| A | | | A continuous layer of small uniform eye-shaped cells bordering the blastocoele |
| B | | | Fewer, larger cells that may not form a continuous layer |
| C | | | Sparse TE cells, may be large |
| D | | | All TE cells degenerate |
| | | | * No ICM or TE score is given for Stage 1 Early Blastocysts<br>** Class 5 consists of embryos which meet the freezing criteria only<br>Freezing/ Biopsy Criteria: Stage 3 or above with a quality score greater than CC<br>(i.e. do NOT freeze or biopsy embryos with a quality score of CC or any embryo with a D (for ICM or TE)<br>**ZP**: Zona Pellucida; **ICM**: Inner cell mass; **TE**: Tropechtoderm |

**Table 1. Blastocyst grading system used by the Massachusetts General Hospital fertility center.** The table shows how the graded embryos were categorized into 5 classes. Classes 1 and 2 primarily consisted of non-blastocysts while classes 3, 4, and 5 consisted of blastocysts. Only embryos belonging to class 5 met the freezing criteria employed the fertility clinic.

The 113 hpi evaluation dataset included images of 2,440 embryos categorized across five classes post-cleaning based on their clinical annotations. Our training set for this classification task used 1,188 images (Class 1: 19.36%; Class 2: 17.68%; Class 3: 20.12%; Class 4: 16.92%; Class 5: 25.92%) with a validation dataset of 510 images (Class 1: 19.41%; Class 2: 18.43%; Class 3: 20.59%; Class 4: 15.69%; Class 5: 25.88%) obtained at 113 hpi. The independent non-overlapping test set consisted of 742 images (Class 1: 19.41%; Class 2: 14.42%; Class 3: 17.38%; Class 4: 11.46%; Class 5: 37.33%). All training was performed within the Keras environment, a popular open-source neural network library designed for python. With the availability of unskewed validation sets prior to augmentation, we used a data generator within Keras for batch generation during training that performed random rotations and flips across all classes on the fly.

### 2.3 Non-Embryoscope image dataset

258 embryo images (Non-blastocysts: 54.65%; Blastocyst: 45.35%) collected through the Society for Reproductive Biologists and Technologists (SRBT) for the Embryo ATLAS project, which were imaged using standard inverted bright-field microscopes annotated by 8 director level embryologists from 8 different fertility practices across the United States, were used for the network evaluation. The threshold for classification was optimized for each architecture but no additional training was performed using the SRBT dataset.

### 2.4 CNN architectures evaluated in this study

Multiple CNN architectures were trained and tested in embryo assessments to identify the best suited network for the task of evaluating embryos. Inception-v3, ResNET, Inception-ResNET, and Xception architectures along with a 40-layer CNN were tested by training them on 113 hpi embryo images for classification. The Inception-v3, ResNET, and Inception-ResNET were trained with the Stochastic Gradient Descent (SGD) optimizer and with learning rates set to 0.0004, 0.0001, and 0.0005, and decay factors of 0.75, 0.5, and 0.5 for every 10 epochs (40 for ResNET), respectively. The 40-layer CNN, similarly, used a learning rate of 0.001 and momentum of 0.5 with an SGD optimizer that had a decay of 0.5 for every 40 epochs. The input size of the embryo image used was $210 \times 210$ pixels and each image were convoluted through 64, 128, 256, and 512 feature maps using 3×3 filters with padding and relu activation. A dropout layer was also used with dropout probability set at 0.5 along with a flattened second-last layer which was connected to the 5-neuron output layer. A few models of ResNET and de-novo CNNs were highlighted here to elucidate the effect of hyperparameters on such networks for the embryo dataset used (Table 2).

6 models for ResNET trained using 113 hpi embryo images are presented here (Table 2). ResNET architecture was used for models 1, 2, 3, and 4, where only model 1 possessed no dropout while the rest (2, 3 and 4) had a layer set at 0.5 probability. Three extra fully connected layers with 1024, 1024, and 512 neurons between the ResNet bottleneck layer and the final classification layer with an additional 20 trainable layers were used for model 5. Model 6 possessed 1024 neurons in a single fully connected layer between the bottleneck layer and the final classification layer, and with a dropout layer set at 0.5. Although only two models are presented here with extra layers, in our overall evaluations adding extra layers to the network did not help the network to learn better. The network was trained by optimizing categorical cross-entropy loss using an SGD optimizer with Nesterov momentum of 0.9 in model 4 and 6 and Adam optimizer in other models. We used learning rates of 0.01 for model 2 and 4, 0.005 for model 3, and 0.001 for all other models. Models with extra layers, dropout and with different optimizers did not help the network to learn better. Model 1 without any extra layers, decay and dropouts performed better than the models evaluated in our study.

For de novo CNN training, 13 models were trained using a 2 layer architecture for models 1 through 8, a 5 layer architecture for models 9, 10, and 11, and a 40 layer architecture for models 12 and 13 (Table 2). The models were tested with different combination of architectural modifications such as batch normalization, dropouts, global average pooling, padding, and dense layers. A learning rate of 0.0005 was used for all presented models except for model 9 that was trained at 0.005. SGD and Adam optimizers were used with and without decay. Even though increasing the number of layers helped in reducing validation loss and improving validation accuracy, the confusion matrices showed that the results were always skewed towards class 5. Therefore, the multi-layer CNNs were not able to learn to classify well between different embryo classes.

The Xception architecture pre-trained with 1.4 million images of ImageNet was used, which performed with a top-1 accuracy of 79% and top-5 accuracy of 94.5% across 1,000 classes of ImageNet database and fine-tuned the pre-training weights across all layers through transfer learning to fit our dataset and differentiate across the categories of embryos by recognizing relevant features. During the transfer learning process, we discarded the last fully connected layer of the original network and added a new fully connected layer, which classifies the features into the defined five categories. The whole network was trained by optimizing categorical cross entropy loss using an SGD Optimizer with Nesterov Momentum of 0.9 and a learning rate of 0.0005 for 113 hpi. The network was trained over 200 epochs and model weights were saved when the lowest validation loss was achieved (early stoppage). The Xception architecture trained over 200 epochs for tasks of evaluating embryo morphologies between the 5 training classes achieved validation loss of 0.8601 and validation accuracy of 63.73%. The dimensions of all embryo images used during training were resized to 210×210 pixels using computer vision libraries (OpenCV).

| Model | Layers between bottleneck and classification layer | Learning rate | Optimizer | Decay | Loss | Accuracy |
|---|---|---|---|---|---|---|
| **ResNET** | | | | | | |
| 1 | Base architecture | 0.001 | Adam | ND | 0.8825 | 0.6011 |
| 2 | Base architecture + dropout (0.5) | 0.01 | SGD | ND | 0.9665 | 0.6137 |
| 3 | Base architecture + dropout (0.5) | 0.005 | Adam | ND | 0.9112 | 0.6039 |
| 4 | Base architecture + dropout (0.5) | 0.01 | SGD | DwS | 0.9269 | 0.6000 |
| 5 | Base architecture + dropout (0.5), additional layers 3 (1024, 1024, 512) + 20 trainable layers | 0.001 | Adam | ND | 1.6235 | 0.2058 |
| 6 | Base architecture + dropout (0.5) + additional layers (1024) | 0.001 | SGD | DwS | 0.9323 | 0.5847 |
| **CNN** | | | | | | |
| 1 | 2 layer3-3 + flatten + 1000 + 5 | 0.0005 | SGD | Decay | 1.4641 | 0.3549 |
| 2 | 2 layer5-5 + flatten + 1000 + 5 | 0.0005 | SGD | Decay | 1.4183 | 0.3843 |
| 3 | 2 layer3-5 + flatten + 1000 + 5 | 0.0005 | SGD | Decay | 1.4198 | 0.3803 |
| 4 | 2 layer5-5 + BN-BN + flatten + 1000 + 5 | 0.0005 | Adam | Decay | 1.3510 | 0.4058 |
| 5 | 2 layer5-5 + BN-BN + global average pooling + 32 + 5 | 0.0005 | Adam | Decay | 1.5231 | 0.3333 |
| 6 | 2 layer5-5 + global average pooling + 32 + 5 | 0.0005 | SGD | Decay | 1.5956 | 0.2588 |
| 7 | 2 layer5-5 + BN-BN + flatten + 1000 + 5 | 0.0005 | SGD | ND | 1.4178 | 0.3647 |
| 8 | 2 layer5-5 + BN-BN + flatten + 1000 + 5 | 0.0005 | Adam | Decay | 1.8889 | 0.3200 |
| 9 | 5 layer5-5 + BN + flatten + 64 + 5 | 0.005 | Adam | Decay | 1.2799 | 0.4313 |
| 10 | 5 layer5-5 + BN + same padding + flatten + 512 + 5 | 0.0005 | SGD | Decay | 1.2255 | 0.4294 |
| 11 | 5 layer5-5 + BN + global average pooling + 16 + 5 | 0.0005 | SGD | Decay | 1.2730 | 0.4215 |
| 12 | 40 layer3-2 + global average pooling + flatten + dense | 0.0005 | SGD | Decay | 1.1581 | 0.4830 |
| 13 | 40 layer3-3 + global average pooling + flatten + dense | 0.0005 | Adam | ND | 1.1689 | 0.5304 |
| **Xception 113 hpi** | | | | | | |
| 1 | Base architecture | 0.0005 | SGD | DwS | 0.8725 | 0.6352 |
| 2 | Base architecture + dropout (0.5) | 0.001 | SGD | DwS | 0.8866 | 0.6333 |
| 3 | Base architecture | 0.0001 | SGD | DwS | 0.9087 | 0.6235 |
| 4 | Base architecture + additional layers 3 (1024,1024,512) + dropout (0.5) | 0.0005 | SGD | DwS | 0.8732 | 0.6549 |
| 5 | Base architecture + dropout (0.5) | 0.0007 | SGD | DwS | 0.8704 | 0.6078 |
| 6 | Base architecture + additional layers (1024) + dropout (0.5) | 0.001 | Adam | ND | 0.8668 | 0.6372 |
| 7 | Base architecture | 0.006 | SGD | ND | 0.8601 | 0.6495 |
| 8 | Base architecture | 0.0008 | SGD | ND | 0.8850 | 0.6196 |

**ND:** No decay; **DwS:** Decay with scheduler; **BN:** Batch Normalization; **SGD:** Stochastic descent gradient

**Table 2.** Models of ResNET, multi-layer CNN, and Xception along with their hyperparameters. The ResNET and multi-layered CNN models were trained using 113 hpi embryo images. Accuracy and loss represent the validation accuracy and loss.

## 2.5 Classification at the inference level

For classification at the inference level, the algorithm outputs five confidence values mapping the probabilities of the tested embryo associated with each of the five training classes. The embryo is categorized into the training class with the highest confidence. Embryos are assigned to the inference classes based on the highest confidence score, which was obtained through the summation of all confidence values associated with the sub-classes of each inference class.

## 2.6 Xception: The effect of hyperparameters

Several trained models of the Xception architecture were evaluated for embryo classifications at 113 hpi and to show the performance related to the choice of learning rate, architectural parameters, and other hyperparameters, we presented 8 models here visualization (Table 2).

We used the vanilla Xception architecture for models 1, 3, 7 and 8, and with dropouts set to 0.5 for models 2 and 5. An extra fully connected layer with 1024 neurons between the Xception bottleneck layer and the final classification layers with dropout was used for model 6. Model 4 possessed three extra fully connected layers with 1024 ,1024, and 512 neurons between the Xception bottleneck layer and the final classification layer and also possessed dropouts. In models 1, 2, 3, 4, 5, and 8 the network was trained by optimizing categorical cross entropy loss using an SGD optimizer with Nesterov momentum of 0.9 with a learning rate of 0.0005, 0.001, 0.0001, 0.0005, 0.0007 and 0.0008, respectively. Models 6 was trained with Adam optimizer with a learning rate of 0.001, while model 7 was trained with SGD and learning rate set at 0.006 with no decay.

## 2.7 Data visualization techniques

Keras vis environment was used for data visualization. Saliency maps were used to visualize the pixels involved in the networks during the decision-making process. We mapped the activations of the activation layer prior to bottleneck. We used the test set images in the generation of the saliency maps. t-distributed stochastic neighbor embedding (t-SNE) was performed to observe the distribution of the test dataset and verify if the CNN was able to isolate embryos into clusters based on their features. We used the fully connected layer after global average pooling which has 2,048 dimensional vectors in visualizing the similarities between the embryo images, as understood by the trained network, using the respective datasets. Initially, a principal component analysis (PCA) was performed to reduce 2,048 dimensions to 50 and then t-SNE was performed to reduce the 50 dimensions into 2 dimensions for visualization. We have utilized PCA for the initial dimensionality reduction to 50 from 2,048 dimensions, since it helps in suppressing noise while improving computational speed (van der Maaten and Hinton, 2008).

# 3. Results

## 3.1 Selection of the optimal neural network

Depending on the complexity of the problem of interest, CNNs generally require large amounts of image data to accurately learn features and differentiate between the categories of classification. However, high-quality medical datasets are scarce and thus, we have transfer learned our networks over ImageNet weights. Proven high-performance CNN architectures such as Inception v3, ResNET, Inception-ResNET-v2, and Xception were retrained using a dataset of 1,188 embryos imaged at 113 hpi and validated using 510 similar images. The same dataset was used to train a 40-layer CNN de novo. All networks were trained with early stoppage rules prioritizing lowest validation loss to minimize overfitting. After training over 200 epochs, the lowest validation loss achieved by these networks were compared. After fine-tuning the hyperparameters for all evaluated networks, the 5-class validation losses of the best models from Inception v3, ResNET, Inception-ResNET-v2, 40-layer CNN, and Xception were 0.9158, 0.8825, 0.8669, 1.1581, and 0.8601, respectively and their 5-class validation accuracies were 60.98%, 60.11%, 59.41%, 48.30%, and 63.73%, respectively (Fig 2, Table 3). Xception architecture achieved the lowest loss for embryo assessments.

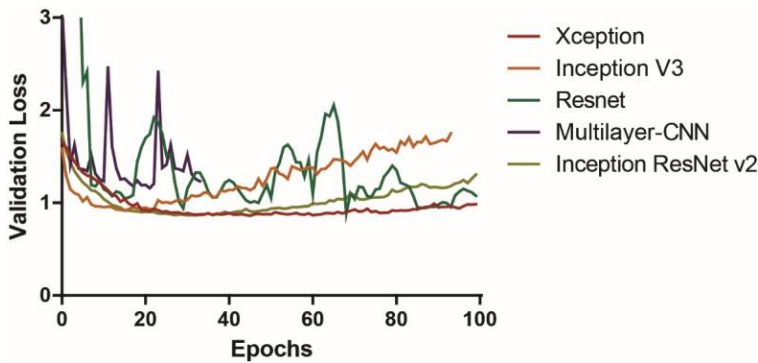

**Fig 2. Comparison of different CNN architectures.** Xception, Inception v3, ResNET, Inception-ResNET v2, and 40-layer CNN were used for embryo classification (5 classes) using 113 hpi embryo images.

| Architectures | Validation losses | Validation accuracies (%) |
|---|---|---|
| Inception v3 | 0.9158 | 60.98 |
| ResNET | 0.8825 | 60.11 |
| Inception-ResNET v2 | 0.8669 | 59.41 |
| 40-layer CNN | 1.1581 | 48.30 |
| Xception | 0.8601 | 64.95 |

**Table 3. Validation losses and accuracies of deep-convolutional neural networks.** Each architecture was transfer learned with a dataset of blastocysts and non-blastocysts imaged at 113 hpi.

Here, we also report 8 models of Xception to demonstrate the effect of different hyperparameters in learning blastocyst data (Fig 3, Table 2). Models benefited from lower learning rates when trained with our dataset. In our tests, SGD performed better for models that evaluated these embryos.

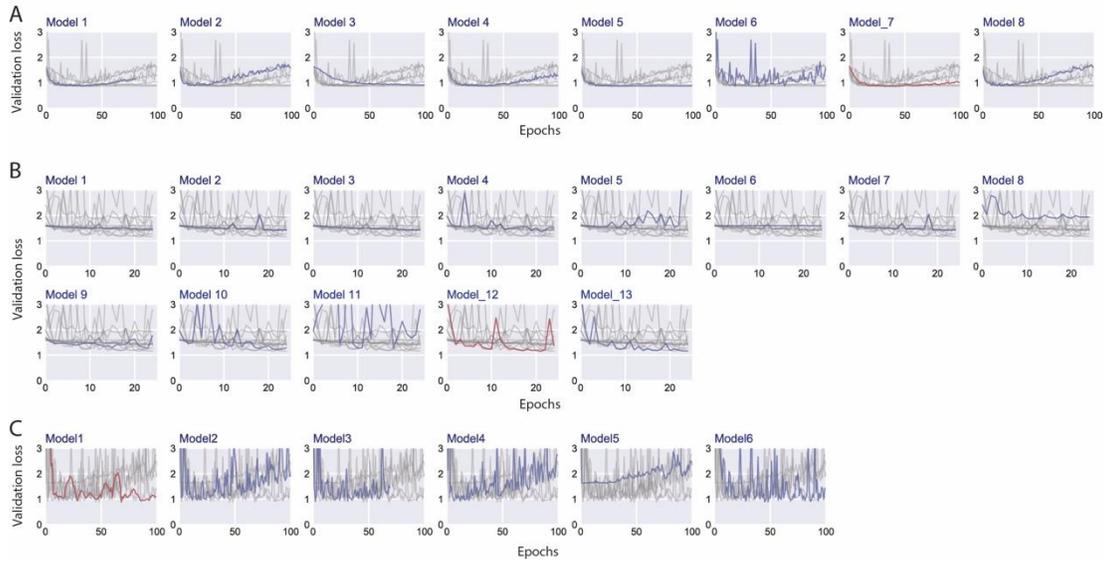

**Fig 3. Comparison of models of ResNET, CNN, and Xception with different hyperparameters.** (A) Validation losses for Xception trained with different hyperparameters using embryo images at 113 hpi. (B) Validation loss of different models of multi-layer CNN with varying hyperparameters trained with 113 hpi embryo images. (C) Validation loss of different models of ResNet to compare the effect of hyperparameters in training embryo images at 113 hpi. The plot with red curve represents the loss curve for the model that achieved the lowest loss among the evaluated models.

We also evaluated simple CNNs designed from scratch, with an increasing number of layers starting from 5 layers to 40 layers (Fig 3, Table 2). Evaluations with multi-layered CNNs indicated that as the complexity of networks increased, better classification performance can be achieved with our embryo dataset as was observed with most of the tested popular neural networks. Interestingly, however, ResNET did not train well regardless of the hyper parameter optimizations employed (Fig 3, Table 2).

## 3.2 Day 5 Embryo Developmental Stage Classification

In clinical practice, transferring high quality blastocyst embryos at the blastocyst stage has been effective in improving embryo selection and thus increasing implantation rates. We therefore evaluated the best performing Xception model with an independent test set of 742 embryos imaged at 113 hpi. The accuracy of the network in categorizing the embryos into two classes of blastocyst and non-blastocyst was 90.97% (CI: 88.67% to 92.93%) (Fig 4). t-SNE visualization and saliency maps showed a clear separation between the two inference groups and reliance on the embryo morphological features by the network for classification (Fig 4, Fig 5), which further indicates that the model makes use of relevant embryo features that are distinct. The highlighted regions by the saliency maps, more specifically, included regions of cellular fragmentation, blastomeres (in cleavage stage embryos/underdeveloped embryos), cavitation, vacuoles, the inner cell mass, and trophectoderm. The confusion matrix of the network in classification between the five training classes (Fig 6) confirmed the model's ability to differentiate between the blastocysts and non-blastocysts and confusions were usually between classes of adjacent quality level.

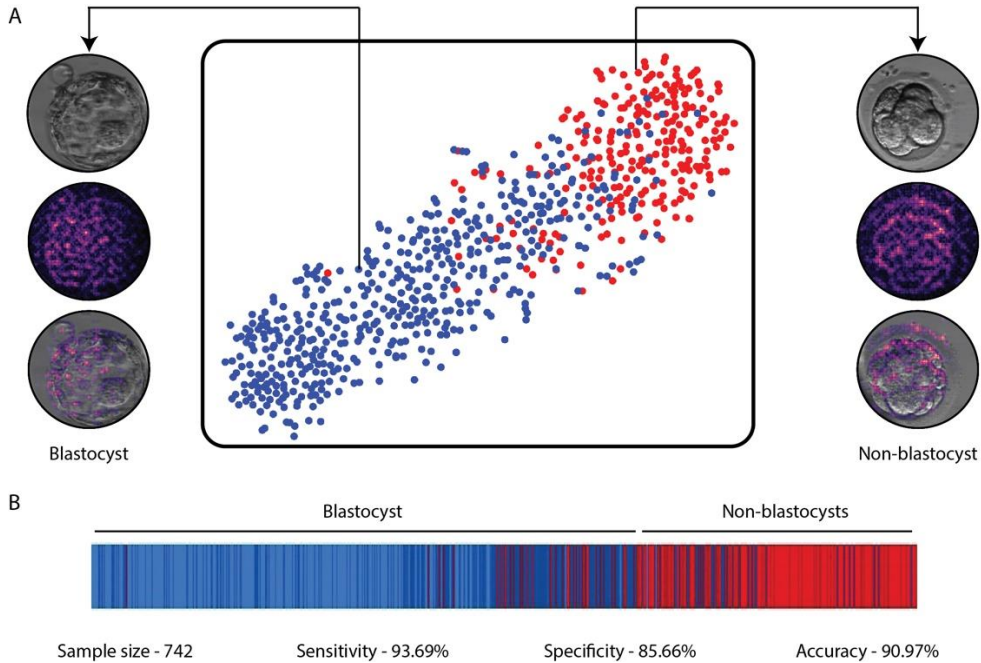

**Fig 4. Evaluation at the blastocyst stage.** (A) The t-SNE plot for the Xception model trained to classify between non-blastocysts (classes 1 and 2) and blastocysts (classes 3, 4, and 5). The saliency map of the two embryos provides an example of the features that network uses to classify embryos on day 5. (B) The composite of bars illustrates the system's performance in evaluating embryos (n=742) from the test set of 97 patients. Each blue bar represents blastocysts and red bar represents non-blastocysts, while the color gradients differentiate the subclasses. The bars are sorted from blastocysts to non-blastocysts (blue-red; classes 5-1) based on their actual labels.

The model's micro-average and macro-average area under the curve (AUC) values at the 5-class training level was calculated to be 0.91 and 0.89 (Fig 7), respectively. The AUC values for the classes 1, 2, 3, 4, and 5 were 0.94, 0.85, 0.88, 0.82, and 0.95, respectively.

The sensitivity and specificity of the Xception model in embryo classification between the two inference classes (blastocysts and non-blastocysts) at 113 hpi were 93.69% (CI: 91.16% to 95.67%) and 85.66% (CI: 80.70% to 89.75%), respectively (n=742 embryos). The positive predictive value (PPV) and negative predictive value (NPV) of the algorithm were 92.74% (CI: 90.42% to 94.54%) and 87.40% (CI: 83.09% to 90.73%), respectively. The AUC metric of the model when evaluated for 2-class classification performance was of 0.963 (CI: 0.947 to 0.975) (Fig 7).

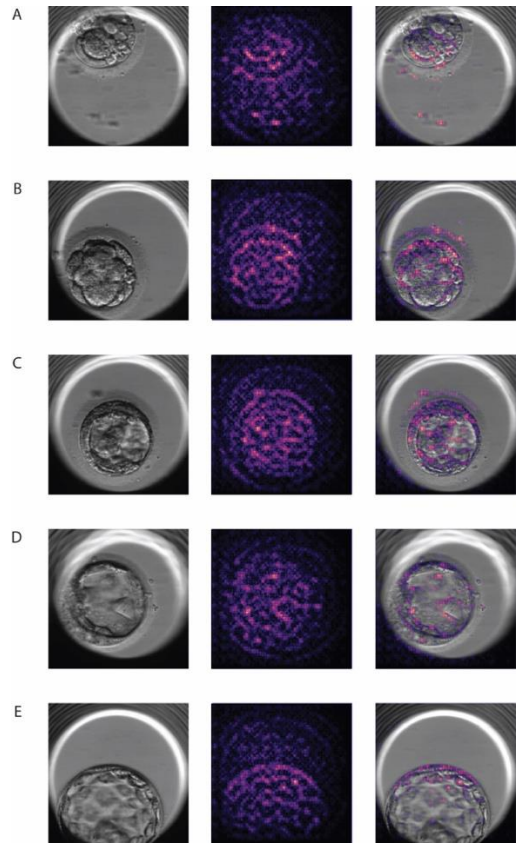

**Fig 5. Saliency maps of embryos assessed at 113 hours post-insemination. The saliency map was extracted from the network to highlight the highest weighted features for the embryo image.** (A) A class 1 embryo category at 113 hpi along with its respective saliency map and saliency map overlaid on the bright-field embryoscope image (B) A class 2 embryo category at 113 hpi along with its respective saliency map and saliency map overlaid on the bright-field embryoscope image. (C) A class 3 embryo category at 113 hpi along with its respective saliency map and saliency map overlaid on the bright-field embryoscope image. (D) A class 4 embryo category at 113 hpi along with its respective saliency map and saliency map overlaid on the bright-field embryoscope image. (E) A class 5 embryo category at 113 hpi along with its respective saliency map and saliency map overlaid on the bright-field embryoscope image. The highlighted regions included regions of cellular fragmentation, blastomeres (in cleavage stage embryos), cavitation, vacuoles, the inner cell mass and trophectoderm.

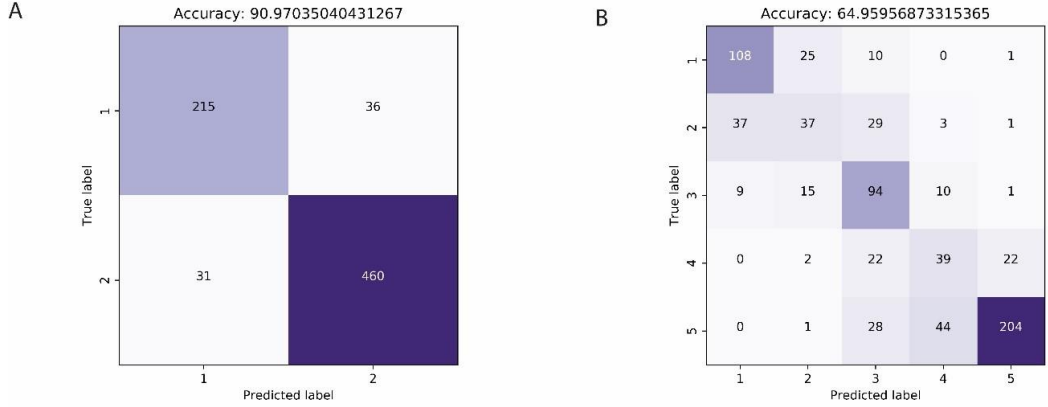

**Fig 6. Confusion Matrices of the neural network in embryo classification tasks.** (A) Confusion matrix for predicting embryos between 2 classes using 113 hpi embryo images. (B) Confusion matrix for classifying embryos between 5 classes using 113 hpi embryo images. Rows represent the historic clinical annotation while the columns indicate the network's predictions.

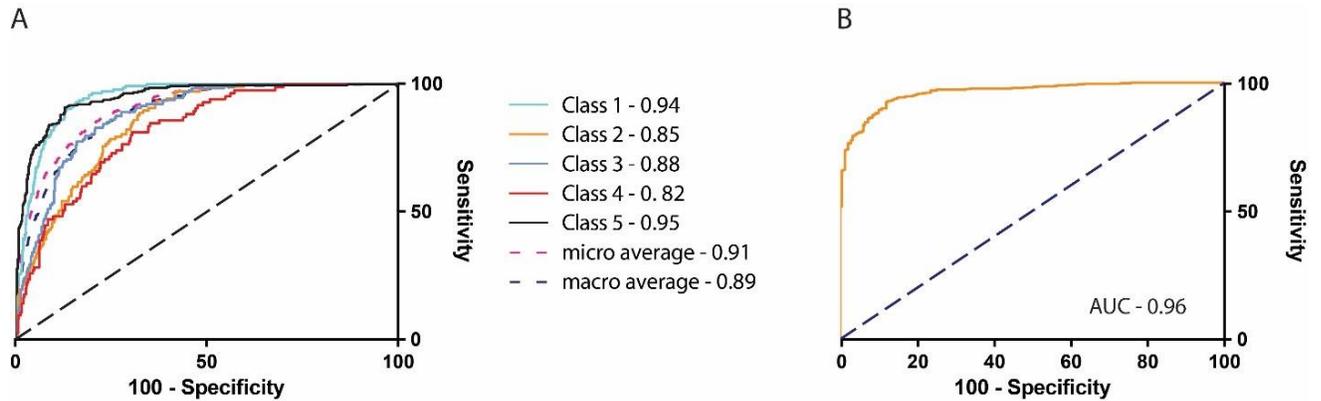

**Fig 7. ROC analysis of the classification task performed by the best Xception model.** (A) ROC analysis performed for all 5 classes of embryo images imaged at 113 hpi. (B) ROC curves for blastocyst and non-blastocyst classification and prediction tasks 113 hpi.

### 3.3 Model performance on domain shifted data

To understand the applicability of the trained model on wider examples of embryo medical datasets we evaluated them using data acquired by different microscopic instruments other than Embryoscope (instrument used for the training dataset), embryo images submitted from 8 fertility practices to the Society for Reproductive Biologists and Technologists (SRBT) for the Embryo ATLAS project (Fig 8), which were imaged using standard inverted bright-field microscopes and annotated by 8 director level embryologists were used. Using a test set of 258 embryo images and without any additional training or image pre-processing, the Xception model performed with an accuracy of 91.47% with a CI of 87.37% to 94.58% in classifying between blastocysts and non-blastocysts (Table 4). The network performed with a sensitivity of 92.31% (CI: 85.90% to 96.42%) and a specificity of 90.78% (CI: 84.75% to 95.00%). The PPV and NPV of the CNN were 89.26% (CI: 83.15% to 93.33%) and 93.43% (CI: 88.34% to 96.39%), respectively with an AUC of 0.975 (CI: 0.948 to 0.990). Saliency maps were also mapped to confirm that the model was using morphological features of the embryos in these images for its classifications. Interestingly, in our evaluations the performance of the Xception model trained on Embryoscope data tested on the domain-shifted SRBT data was similar to its original performance. However, all other networks showed a drastic drop in performance in comparison to their original performance, when tested on the domain-shifted SRBT dataset (Table 4).

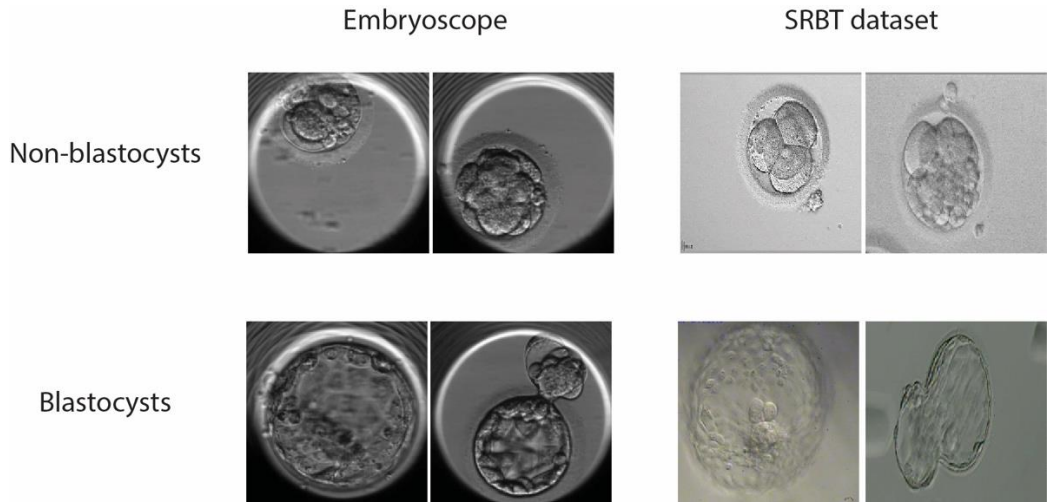

**Fig 8. Representative images of Embryoscope and SRBT datasets.**

Images collected using the Embryoscope system was used in training the network and the SRBT images collected at different clinical laboratories using inverted-brightfield microscopes were used for our experiments with domain-shifted data. All images here are resized to 210x210 pixels to reflect the input of the neural networks.

|  | Test set accuracies (%) | |
| :---: | :---: | :---: |
| **Architecture** | **Embryoscope (n=742)** | **SRBT (n=258)** |
| Xception | 90.97 | 91.47 |
| Inception v3 | 90.70 | 83.29 |
| ResNET | 88.27 | 74.81 |
| Inception ResNET | 90.97 | 84.88 |
| Multilayer CNN | 81.11 | 60.85 |

**Table 4. Performance of different architectures on embryoscope and SRBT datasets.** All models tested were optimized through tuning their hyperparameters.

## 4. Discussion and Conclusions

Here, we report the development and evaluation of an AI-based approach for automated human embryo assessment of embryo development by 113 hpi. In the recent years, due to the upsurge in deep-learning research, various complex neural network architectures have been proposed and used for image recognition tasks and performance of these architectures is highly dependent on the task. In our study, CNN approaches were evaluated using a dataset of 113 hpi images in classifying embryos based on their morphological quality. Firstly, our evaluations of whether a simple deep CNN, of up to 40 layers, was sufficient for efficiently assessing embryos indicated that more sophisticated networks may be preferable than simply stacked deeper networks. Inception, (48 layers) which uses inception modules that are composed of multiple filters of different sizes, over simple convolution layers showed significantly better performance than our de novo CNNs. We also evaluated the performance of other popular networks such as ResNET, which originally introduced residual blocks, and the hybrid Inception-ResNET. However, it was surprising that ResNet performed with poor accuracy and learning regardless of hyper-parameter optimizations employed in our evaluations. Xception was developed with Inception as its base architecture, while adding residual blocks and replacing the convolutions in original inception modules with depthwise seperable convolutions.

Our observations have shown that Xception performed best in learning the categorical embryo data and was able classify them based on their morphological quality. Most machine-learning approaches reported in the literature for embryo evaluations are usually not capable of generalization across instruments due to their relatively rigid structure of data collection and analysis or due to domain specific learning. Xception, interestingly, performed well on domain shifted data (SRBT dataset) that was acquired through different imaging systems, which is uncommon with machine-learning- or computer vision-based approaches (Tommasi et al., 2015). Saliency maps of embryos imaged at the blastocyst stage (113 hpi), highlight the whole embryo as regions of interest which is further indicative of well-trained model that utilizes features that are relevant for classification (Fig 5). However, while the study is suggestive that the thus trained Xception model more robust even with domain shifted data, additional evaluations are required to be conclusive. It is, however, encouraging for future studies that make use of Xception for embryo morphological analyses.

The primary goal in an IVF procedure is to culture and transfer an embryo that will result in a healthy baby. Embryologists, therefore, try to identify the embryo of highest quality for transfer and to avoid embryos of the lower quality from a cohort for patients with good prognosis. A neural network's raw ability of separating embryos between five classes does not directly benefit such clinical processes. Furthermore, the 5-class classification accuracy should be taken with caution since it also affected by the annotating embryologist's ability to repeatably and consistently categorize embryos based on their morphology, which has been observed to be not ideal. However, the evaluations are useful in understanding the network learning. These networks, to be clinically viable, need to be modified to suit the applications. For example, Xception correctly classified >99.5% of the highest quality blastocysts as good embryos (blastocysts) which is of critical importance, clinically, when identifying embryos suited for transfer. In this work, primarily to minimize sparsity of data in limited dataset, we have classified embryos based on hierarchical classification system that consolidates the MGH blastocyst categorization into 5 classes though embryologists have highlighted that 5-class system for embryo morphology-based classification may be more beneficial over commonly used 3 class classification. For the study, we have consolidated our network's 5-class output to 2 inference classes and differentiated embryos between blastocysts and non-blastocysts to highlight its performance on a more universal classification system (blastocysts and non-blastocysts) since embryo categorization criteria tends to vary with each clinic.

The work presented here is an example of demonstrating how deep-learning techniques can be used in medicine particularly in an IVF procedure. The IVF community can greatly benefit from the modern advancements in machine learning. Our future work will be focused on many valuable applications and goals, such as predicting embryo developmental outcomes at earlier timepoints, studying the use of neural networks in aiding routine clinical tasks, and in predicting the eventual outcome of embryos. Our presented results show the promise of using neural networks in accurate embryo assessments with the potential to eventually improve IVF practices in both resource-rich and resource-poor settings regardless of the center's experience and infrastructure.

## Acknowledgements

The authors would like to thank embryology staff from Massachusetts General Hospital for participating in this study. This work was supported by the Brigham and Women's Hospital (BWH) Precision Medicine Developmental Award (BWH Precision Medicine Program) and Partners Innovation Discovery Grant (Partners Healthcare). It was also partially supported through 1R01AI118502, and R01AI138800 Awards (National Institute of Health). The funders had no role in study design, data collection and analysis, decision to publish, or preparation of the manuscript.

**Author contributions**:

**Prudhvi Thirumalaraju:** Conceptualization, Methodology, Software, Validation, Investigation, Writing – Review & Editing, Visualization. **Manoj Kumar Kanakasabapathy:** Conceptualization, Methodology, Validation, Formal analysis, Investigation, Writing - Original Draft, Supervision, Project administration. **Charles L Bormann:** Conceptualization, Methodology, Investigation, Data Curation, Writing - Review & Editing, Supervision. **Raghav Gupta:** Methodology, Software, Visualization, Writing - Review & Editing. **Rohan Pooniwala:** Methodology, Software, Visualization, Writing - Review & Editing. **Hemanth Kandula:** Methodology, Software, Visualization, Writing - Review & Editing. **Irene Souter:** Data Curation, Resources, Writing - Review & Editing. **Irene Dimitriadis:** Data Curation, Resources, Writing - Review & Editing. **Hadi Shafiee:** Conceptualization, Validation, Resources, Writing - Review & Editing, Supervision, Project administration, Funding acquisition

**Data and materials availability**: Restrictions apply to the availability of the medical training/validation data, which were used with permission for the current study, and so are not publicly available. Some data may be available from the authors upon reasonable request and with permission of the Massachusetts General Hospital.

# References

Barash, O., Ivani, K., Huen, N., Willman, S., Weckstein, L., 2017. Morphology of the blastocysts is the single most important factor affecting clinical pregnancy rates in IVF PGS cycles with single embryo transfers. Fertil. Steril. 108, e99.

Birenbaum-Carmeli, D., 2004. 'Cheaper than a newcomer': on the social production of IVF policy in Israel. Sociol Health Illn 26, 897-924. CDC, 2015. Fertility Clinic Success Rates Report.

Chollet, F., 2016. Xception: Deep Learning with Depthwise Separable Convolutions. arXiv, 1610.02357.

Conaghan, J., Chen, A.A., Willman, S.P., Ivani, K., Chenette, P.E., Boostanfar, R., Baker, V.L., Adamson, G.D., Abusief, M.E., Gvakharia, M., Loewke, K.E., Shen, S., 2013. Improving embryo selection using a computer-automated time-lapse image analysis test plus day 3 morphology: results from a prospective multicenter trial. Fertil Steril 100, 412-419.e415.

Demko, Z.P., Simon, A.L., McCoy, R.C., Petrov, D.A., Rabinowitz, M., 2016. Effects of maternal age on euploidy rates in a large cohort of embryos analyzed with 24-chromosome single-nucleotide polymorphism–based preimplantation genetic screening. Fertil. Steril. 105, 1307-1313.

Dimitriadis, I., Bormann, C.L., Kanakasabapathy, M.K., Thirumalaraju, P., Gupta, R., Pooniwala, R., Souter, I., Rice, S.T., Bhowmick, P., Shafiee, H., 2019a. Deep convolutional neural networks (CNN) for assessment and selection of normally fertilized human embryos. Fertil. Steril. 112, e272.

Dimitriadis, I., Bormann, C.L., Thirumalaraju, P., Kanakasabapathy, M., Gupta, R., Pooniwala, R., Souter, I., Hsu, J.Y., Rice, S.T., Bhowmick, P., Shafiee, H., 2019b. Artificial intelligence-enabled system for embryo classification and selection based on image analysis. Fertil. Steril. 111, e21.

Einarsson, S., Bergh, C., Friberg, B., Pinborg, A., Klajnbard, A., Karlström, P.-O., Kluge, L., Larsson, I., Loft, A., Mikkelsen-Englund, A.-L., Stenlöf, K., Wistrand, A., Thurin-Kjellberg, A., 2017. Weight reduction intervention for obese infertile women prior to IVF: a randomized controlled trial. Hum. Reprod. 32, 1621-1630.

Erenus, M., Zouves, C., Rajamahendran, P., Leung, S., Fluker, M., Gomel, V., 1991. The effect of embryo quality on subsequent pregnancy rates after in vitro fertilization. Fertil. Steril. 56, 707-710.

Filho, E.S., Noble, J.A., Wells, D., 2010. A Review on Automatic Analysis of Human Embryo Microscope Images. Open Biomed Eng J 4, 170-177.

Hariton, E., Dimitriadis, I., Kanakasabapathy, M.K., Thirumalaraju, P., Gupta, R., Pooniwala, R., Souter, I., Rice, S.T., Bhowmick, P., Ramirez, L.B., Curchoe, C.L., Swain, J.E., Boehnlein,

L.M., Bormann, C.L., Shafiee, H., 2019. A deep learning framework outperforms embryologists in selecting day 5 euploid blastocysts with the highest implantation potential. Fertil. Steril. 112, e77-e78.

He, K., Zhang, X., Ren, S., Sun, J., 2015. Deep Residual Learning for Image Recognition, ArXiv e-prints.

Hill, G.A., Freeman, M., Bastias, M.C., Jane Rogers, B., Herbert, C.M., III, Osteen, K.G., Wentz, A.C., 1989. The influence of oocyte maturity and embryo quality on pregnancy rate in a program for in vitro fertilization-embryo transfer Fertil. Steril. 52, 801-806.

Kanakasabapathy, M., Dimitriadis, I., Thirumalaraju, P., Bormann, C.L., Souter, I., Hsu, J., Thatcher, M.L., Veiga, C., Shafiee, H., 2019a. An inexpensive, automated artificial intelligence (AI) system for human embryo morphology evaluation and transfer selection. Fertil. Steril. 111, e11.

Kanakasabapathy, M.K., Thirumalaraju, P., Bormann, C.L., Kandula, H., Dimitriadis, I., Souter, I., Yogesh, V., Kota Sai Pavan, S., Yarravarapu, D., Gupta, R., Pooniwala, R., Shafiee, H., 2019b. Development and evaluation of inexpensive automated deep learning-based imaging systems for embryology. Lab Chip 19, 4139-4145.

Kanakasabapathy, M.K., Thirumalaraju, P., Gupta, R., Pooniwala, R., Kandula, H., Souter, I., Dimitriadis, I., Bormann, C.L., Shafiee, H., 2019c. Improved monitoring of human embryo culture conditions using a deep learning-derived key performance indicator (KPI). Fertil. Steril. 112, e70-e71.

Khosravi, P., Kazemi, E., Zhan, Q., Malmsten, J.E., Toschi, M., Zisimopoulos, P., Sigaras, A., Lavery, S., Cooper, L.A.D., Hickman, C., Meseguer, M., Rosenwaks, Z., Elemento, O., Zaninovic, N., Hajirasouliha, I., 2019. Deep learning enables robust assessment and selection of human blastocysts after in vitro fertilization. NPJ Digit Med 2, 21.

Machtinger, R., Racowsky, C., 2013. Morphological systems of human embryo assessment and clinical evidence. Reprod. Biomed. Online 26, 210-221.

Mascarenhas, M.N., Flaxman, S.R., Boerma, T., Vanderpoel, S., Stevens, G.A., 2012. National, Regional, and Global Trends in Infertility Prevalence Since 1990: A Systematic Analysis of 277 Health Surveys. PLoS Med. 9, e1001356.

Matos, F.D., Rocha, J.C., Nogueira, M.F.G., 2014. A method using artificial neural networks to morphologically assess mouse blastocyst quality. Journal of Animal Science and Technology 56, 15.

Osman, A., Alsomait, H., Seshadri, S., El-Toukhy, T., Khalaf, Y., 2015. The effect of sperm DNA fragmentation on live birth rate after IVF or ICSI: a systematic review and meta-analysis. Reprod. Biomed. Online 30, 120-127.

Paulson, R.J., Sauer, M.V., Lobo, R.A., 1990. Embryo implantation after human in vitro fertilization: importance of endometrial receptivity. Fertil. Steril. 53, 870-874.

Racowsky, C., Kovacs, P., Martins, W.P., 2015. A critical appraisal of time-lapse imaging for embryo selection: where are we and where do we need to go? J. Assist. Reprod. Genet. 32, 1025-1030.

Rocha, J.C., Passalia, F.J., Matos, F.D., Takahashi, M.B., Ciniciato, D.d.S., Maserati, M.P., Alves, M.F., de Almeida, T.G., Cardoso, B.L., Basso, A.C., Nogueira, M.F.G., 2017a. A Method Based on Artificial Intelligence To Fully Automatize The Evaluation of Bovine Blastocyst Images. Sci. Rep. 7, 7659.

Rocha, J.C., Passalia, F.J., Matos, F.D., Takahashi, M.B., Maserati, M.P., Jr., Alves, M.F., de Almeida, T.G., Cardoso, B.L., Basso, A.C., Nogueira, M.F.G., 2017b. Automatized image processing of bovine blastocysts produced in vitro for quantitative variable determination. Sci Data 4, 170192.

Szegedy, C., Ioffe, S., Vanhoucke, V., Alemi, A., 2016. Inception-v4, Inception-ResNet and the Impact of Residual Connections on Learning, ArXiv e-prints.

Szegedy, C., Vanhoucke, V., Ioffe, S., Shlens, J., Wojna, Z., 2015. Rethinking the Inception Architecture for Computer Vision, ArXiv e-prints.

Thirumalaraju, P., Bormann, C.L., Kanakasabapathy, M.K., Kandula, H., Shafiee, H., 2019a. Deep learning-enabled prediction of fertilization based on oocyte morphological quality. Fertil. Steril. 112, e275.

Thirumalaraju, P., Hsu, J.Y., Bormann, C.L., Kanakasabapathy, M., Souter, I., Dimitriadis, I., Dickinson, K.A., Pooniwala, R., Gupta, R., Yogesh, V., Shafiee, H., 2019b. Deep learning-enabled blastocyst prediction system for cleavage stage embryo selection. Fertil. Steril. 111, e29.

Thirumalaraju, P., Kanakasabapathy, M.K., Gupta, R., Pooniwala, R., Kandula, H., Souter, I., Dimitriadis, I., Bormann, C.L., Shafiee, H., 2019c. Automated quality assessment of individual embryologists performing ICSI using deep learning-enabled fertilization and embryo grading technology. Fertil. Steril. 112, e71.

Tommasi, T., Patricia, N., Caputo, B., Tuytelaars, T., 2015. A Deeper Look at Dataset Bias, arXiv e-prints.

Toner, J.P., 2002. Progress we can be proud of: U.S. trends in assisted reproduction over the first 20 years. Fertil Steril 78, 943-950.

Tran, D., Cooke, S., Illingworth, P.J., Gardner, D.K., 2019. Deep learning as a predictive tool for fetal heart pregnancy following time-lapse incubation and blastocyst transfer. Hum. Reprod. 34, 1011-1018.

Turchi, P., 2015. Prevalence, Definition, and Classification of Infertility, In: Cavallini, G., Beretta, G. (Eds.), Clinical Management of Male Infertility. Springer International Publishing, Cham, pp. 5-11.

Vaegter, K.K., Lakic, T.G., Olovsson, M., Berglund, L., Brodin, T., Holte, J., 2017. Which factors are most predictive for live birth after in vitro fertilization and intracytoplasmic sperm injection (IVF/ICSI) treatments? Analysis of 100 prospectively recorded variables in 8,400 IVF/ICSI single-embryo transfers. Fertil. Steril. 107, 641-648.e642.

van der Maaten, L., Hinton, G., 2008. Visualizing Data using t-SNE. Journal of Machine Learning Research 9, 2579-2605.

Wong, C., Chen, A.A., Behr, B., Shen, S., 2013. Time-lapse microscopy and image analysis in basic and clinical embryo development research. Reprod. Biomed. Online 26, 120-129.